\documentclass[12pt,a4paper]{article}
\usepackage{amsmath,amssymb,epsfig}
\newcommand{\beq}{\begin{eqnarray}}
\newcommand{\eeq}{\end{eqnarray}}

\newcommand{\be}{\begin{equation}}
\newcommand{\ee}{\end{equation}}

\newcommand{\bm}{\begin{multline}}
\newcommand{\fm}{\end{multline}}

\begin{document}
\numberwithin{equation}{section}
\setlength{\unitlength}{.8mm}

\begin{titlepage} 
\vspace*{0.5cm}
\begin{center}
{\Large\bf The finite size spectrum of the 2-dimensional O$(3)$
nonlinear $\sigma$-model}
\end{center}
\vspace{2.5cm}
\begin{center}
{\large J\'anos Balog and \'Arp\'ad Heged\H us}
\end{center}
\bigskip
\begin{center}
Research Institute for Particle and Nuclear Physics,\\
Hungarian Academy of Sciences,\\
H-1525 Budapest 114, P.O.B. 49, Hungary\\ 
\end{center}
\vspace{3.2cm}
\begin{abstract}\noindent

Nonlinear integral equations are proposed for the description
of the full finite size spectrum of the 2-dimensional O$(3)$ nonlinear 
$\sigma$-model in a periodic box. Numerical results for the energy
eigenvalues are compared to the rotator spectrum and perturbation
theory for small volumes and with the recently proposed generalized
L\"uscher formulas at large volumes.
\end{abstract}

\end{titlepage}


\newsavebox{\Sau}
\sbox{\Sau}{\begin{picture}(140,25) (-70,-12.5)

\put(15,0){\circle{3}}
\put(5,0){\circle{3}}
\put(-5,0){\circle{3}}
\put(-15,0){\circle{3}}
\put(-25,10){\circle{3}}
\put(-25,-10){\circle*{3}}

\put(-3.5,0){\line(1,0){7}}
\put(-13.5,0){\line(1,0){7}}
\put(6.5,0){\line(1,0){7}}

\put(-16.1,1.1){\line(-1,1){7.7}}
\put(-16.1,-1.1){\line(-1,-1){7.7}}

\multiput(17.5,0) (1,0) {6} {\circle*{0.2}}

\put(-21,-11){\makebox(0,0)[t]{{\protect\scriptsize 0}}}
\put(-21,11){\makebox(0,0)[t]{{\protect\scriptsize 1}}}
\put(-12.5,-3){\makebox(0,0)[t]{{\protect\scriptsize 2}}}
\put(-3,-3){\makebox(0,0)[t]{{\protect\scriptsize 3}}}
\put(5.5,-3){\makebox(0,0)[t]{{\protect\scriptsize 4}}}
\put(14,-3){\makebox(0,0)[t]{{\protect\scriptsize 5}}}
\end{picture}}


\section{Introduction}

The study of finite size (FS) effects in quantum field theories has recently
been received a lot of attention. While understanding the structure
of FS effects has always been an important part of the theory of quantum
fields (in particular in the numerical simulation of lattice field theories),
this renewed interest is due to the important role FS effects are playing
in the verification of the AdS/CFT correspondence \cite{AdS}. Partially 
motivated by similarity to the AdS/CFT problem, nonlinear integral equations
(NLIE) have been proposed \cite{KaziO4} to describe the spectrum of
excited states in the 2-dimensional O$(4)$ nonlinear $\sigma$-model
confined to a finite, periodic box. Motivated also by a problem in
the AdS/CFT correspondence, a generalization of L\"uscher's 
formulas \cite{Luscher}, giving the leading large volume dependence for 
all excited states were proposed \cite{BJ}. This result was succsessfully
applied for calculating the 4-loop \cite{BJ} and 5-loop \cite{BHJL}
anomalous dimension of an important operator in the AdS/CFT correspondence.

In this paper we propose a set of nonlinear integral equations and
corresponding quantization conditions for the description of the
full finite size spectrum of the O$(3)$ NLS model.
Based on the proposal that the $O(3)$ NLS model can be represented as a limit of appropriately
perturbed $Z_N$ parafermion conformal field theories (CFT), 
the Thermodynamic Bethe Ansatz (TBA) equations for the ground state of the $O(3)$ NLS model 
were proposed first\footnote{Later a generalization to a more general class
of $G/H$ coset models were given in \cite{PF}.}
in \cite{FZ}. The equations could be formulated 
in a rather elegant form in terms of the incidence 
matrix  $I_{jk}$ of an infinite Dynkin diagram of type $\cal D$ depicted in 
Figure 1:
\begin{equation} \label{TBA}
\log y_k(\theta)=-\ell \delta_{k0}
\, \cosh \theta + \sum\limits_{j=0}^{\infty} I_{kj} \,
 (K*\ln Y_j)(\theta), \quad
Y_k(\theta)=1+y_k(\theta),  \quad k=0,1,2,...
\end{equation}
where $\ell$ is the volume\footnote{We measure all energies and lengths
in units defined by the infinite volume mass gap of the model.},
$K(\theta)$ denotes the TBA kernel $K(\theta)=
\frac{1}{2\pi\cosh(\theta)}$  and $*$ denotes
convolution: $(f*g)(x)=\int\limits_{-\infty}^{\infty} 
{\rm d}y \, f(x-y) g(y).$
\begin{figure}[htbp]
\begin{center}
\begin{picture}(280,30)(-140,-15)
\put(-114,-7) {\usebox{\Sau}}
\put(-130,-15){\parbox{130mm}{\caption{ \label{6f}\protect 
{\footnotesize
Dynkin-diagram associated with the Y-system of the  $O(3)$ $\sigma$-model. }}}}
\end{picture}
\end{center}
\end{figure}
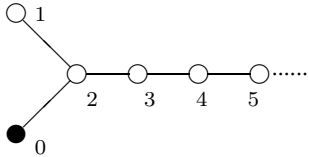
The energy of the ground state can be expressed in terms of the solution of (\ref{TBA}) by a simple
integral expression:
\begin{equation} \label{TBAenergy}
E= -\int\limits_{-\infty}^{\infty} \frac{{\rm d}\theta}{2 \pi}
\, \cosh \theta \, \ln Y_0(\theta).
\end{equation}
It was also shown that the solutions of the ground state TBA
equations satisfy certain functional equations, the so-called Y-system equations, which take the 
following form for the $O(3)$ NLS model:
\begin{equation} \label{Y}
y_k(\theta+i \, q) \, y_k(\theta-i \, q)=
\prod\limits_{j=0}^{\infty} Y_{j}(\theta)^{I_{kj}}, \qquad 
q=\frac{\pi}{2} \qquad k=0,1,....  
\end{equation}
Detailed studies of analogous problems for
integrable lattice models \cite{KPS} indicated that the Y-system is 
universal in the sense that
different solutions of the same Y-system describe 
all excited states of the model. 
The difference between the  solutions is encoded 
into their analytic structure. Thus 
the excited states can also be described by TBA equations, which are 
similar to the ground state TBA
equations but are supplemented by additional source terms and 
quantization conditions.

For relativistic quantum field theories (QFT) this phenomenon was first 
observed in~\cite{DT}, where for certain
perturbed CFTs it was demonstrated that excited state TBA equations 
can be obtained by the analytic
continuation of some parameters of the ground state equations. Later,
based on an integrable lattice regularization, universality of 
the Y-system for all excited states was 
demonstrated and excited state TBA equations were derived in the sine-Gordon 
model \cite{BH-SG}. In this model it was also shown
that the source terms for excitations can be determined by
studying the large volume asymptotics of the Y-functions.
 
The above examples suggested that it is true in general also for 
relativistic QFTs that the same Y-system
describes all excited states of the model and the TBA equations 
for the ground state and excited states
differ only by source terms and additional quantization conditions, 
which can be determined from the
knowledge of the infrared (IR) asymptotics of the Y-functions.    
Accepting the above hypothesis 1-particle TBA equations were proposed 
for the $O(3)$ and $O(4)$
NLS models by an appropriate modification of the analytic properties of the 
vacuum Y-functions \cite{BH34}. 

The disadvantage of the TBA approach to sigma models is 
that the number of unknown functions is infinite. It is desirable
to find an equivalent nonlinear integral equation (NLIE)
description of the problem, with only a few unknown functions. 
Although such a set of NLIEs is equivalent to the Y-system and TBA,
it is a very useful technical tool that makes the analysis of
finite size effects  more efficient. 
In the cases of interest a Y-system is equivalent to a T-system and the latter 
can be solved by an auxiliary linear problem called TQ-relations. 
As it was shown in a series of papers the TQ-relations
are the basic tool for the derivation of an NLIE \cite{KPnlie}. 
In \cite{DdV-SG,BLZ,SSG} integrable lattice regularizations were used
to derive NLIEs for certain QFTs, while in the case of the $O(4)$ NLS model 
\cite{KaziO4} started directly from the TQ-relations of the continuum theory.

In a series of integrable lattice models \cite{KPnlie} and relativistic 
QFTs \cite{DdV-SG,BLZ,SSG} it has also been demonstrated
that similarly to the TBA case the NLIE for the ground state and 
excited states differ only in source 
terms and quantization conditions, which can be determined from the IR 
analysis of the unknown functions.
Using this hypothesis and starting from the ground state equations 
\cite{Dunning,HegO4} 1-particle 
NLIEs were proposed for the $O(3)$ and $O(4)$ NLS models \cite{HegO34}.   

The organization of the paper is as follows. In the next section we propose
a set of nonlinear integral equations and quantization conditions for the
full spectrum of the periodic O$(3)$ NLS model. In Section 3 we study the
large volume limit and give a detailed analysis of the spectrum of 
2-particle states in this limit using the solution of the asymptotic
Bethe Ansatz. In Section 4 we discuss our numerical results, both for small 
volumes, where comparison to the rotator spectrum and perturbation theory, 
and for large volumes, where comparison with the Bajnok-Janik formulas, 
are presented. Technical details are explained in Appendices~A~and~B.

\section{The nonlinear integral equations}

Our starting point in this paper is the hypothesis that the same Y-system 
describes all excited states of the $O(3)$ NLS model.
We generalize the NLIE technique to describe the complete finite size 
spectrum of the model. The generalization is based on the equivalent 
formulations of the Y-system (\ref{Y}) through a T-system and
TQ-relations. With the help of the TQ-relations we can define appropriate 
auxiliary functions in terms of which the infinite set of TBA equations 
can be reduced to a finite set of equations, 
containing only a few unknown functions. 

In the large $\ell$ limit eigenstates of the model are multi-particle 
states satisfying the Bethe-Yang equations, which provide quantization
conditions on the set of particle rapidities, $X=\{\theta_k\}$,
$k=1,2,\dots,N$. This is constructed as follows. We take an $M$-magnon
solution of the asymptotic Bethe equations:
\begin{equation} \label{uj}
\frac{\varphi_0(u_j+i \, \pi)}{\varphi_0(u_j-i\pi)}
=-\frac{Q_0(u_j+i\, \pi)}{Q_0(u_j-i \, \pi)}, \qquad 
j=1,..,M,   
\end{equation}
where
\begin{equation} \label{fi0Q}
\varphi_0(\theta)=\prod\limits_{k=1}^{N}(\theta-\theta_k), 
\qquad Q_0(\theta)=\prod\limits_{j=1}^M 
(\theta-u_j),
\end{equation}
and the $u_j$s are the magnon rapidities and their number $M\leq N$ 
determines the isospin of the $N$-particle state through the formula $J=N-M.$ 

Next we construct the T-system elements for $k=0,1,\dots$
\begin{equation} 
T_k(\theta)=Q_0(\theta+i(k+1)q)Q_0(\theta-i(k+1)q)
\sum_{j=0}^k \xi(\theta+i(k-2j)q),
\label{asyT}
\end{equation}
where
\begin{equation}
\xi(\theta)=\frac{\varphi_0(\theta+iq)\varphi_0(\theta-iq)}
{Q_0(\theta+iq)Q_0(\theta-iq)}.
\end{equation}
In particular,
\begin{equation}
T_0(\theta)=\varphi_0(\theta+iq)\,\varphi_0(\theta-iq),\qquad
T_1(\theta)=\varphi_0(\theta)\,\tilde T_1(\theta),
\label{T0T1}
\end{equation}
where
\begin{equation}
\tilde T_1(\theta)=\frac{1}{Q_0(\theta)}\left\{
\varphi_0(\theta+i\pi)\,Q_0(\theta-i\pi)+
\varphi_0(\theta-i\pi)\,Q_0(\theta+i\pi)\right\},
\label{tilT1}
\end{equation}
which is a polynomial of degree $N$ as a consequence of the
Bethe Ansatz equations (\ref{uj}). It can be shown also that all $T_k$
are polynomials of degree $2N$.

Using the definition
\begin{equation}
\phi(\theta)=\varphi_0(\theta)\,\varphi_0(\theta+2iq)
\label{phi}
\end{equation}
it is possible to verify that the functions $\phi,T_0,T_1,\dots$ form 
a semi-infinite T-system
\begin{equation} \label{T}
T_k(\theta+i \, q) \, T_k(\theta-i \, q)-T_{k-1}(\theta)\, 
T_{k+1}(\theta)=\phi(\theta+i \, k\, q) \,
\bar{\phi}(\theta-i \, k \, q ),
\end{equation}
where $\bar{\phi}$ is the complex conjugate\footnote{Throughout
this paper $\bar f$ denotes the complex conjugate of the function $f$.}
of $\phi$. The system is semi-infinite because it can be consistently
truncated putting $T_{-1}(\theta)\equiv0$.

Next one constructs the Y-system elements
\begin{equation} \label{yT}
y_k(\theta)=\frac{T_{k-1}(\theta) \, T_{k+1}(\theta)}{\phi(\theta+i\, 
k \, q) \, \bar{\phi}(x-i \, k\, q)},
\end{equation}
\begin{equation} \label{YT}
Y_k(\theta)=\frac{T_{k}(\theta+ i \, q) \, T_{k}(\theta-i \, q)}
{\phi(\theta+i\, k \, q) \, \bar{\phi}(x-i \, k\, q)},  
\end{equation}
which satisfy the Y-system equations
\begin{equation}
y_k(\theta+iq)\,y_k(\theta-iq)=Y_{k+1}(\theta)\,Y_{k-1}(\theta)
\label{Ysys}
\end{equation}
for $k=2,3,\dots$ and
\begin{equation}
y_1(\theta+iq)\,y_1(\theta-iq)=Y_2(\theta).
\label{Ysys1}
\end{equation}

The main result in this construction is that
\begin{equation}
y_1(\theta)=\Lambda(\theta+iq\vert\underline{\theta}),
\end{equation}
where $\Lambda(\theta,\underline\theta)$
denotes the eigenvalue of the transfer matrix made out of the
two-particle S-matrix  $\hat{S}_{ij}(\theta)$ of the $O(3)$ NLS model:
\begin{equation} \label{TO3}
T(\theta,\underline\theta)=\mbox{Tr}_0 (\hat{S}_{01}(\theta-\theta_1) 
\,\hat{S}_{02}(\theta-\theta_2)...\hat{S}_{0N}(\theta-\theta_N)).
\end{equation}
More precisely, here we deal with the spin-1 transfer matrix, where
both in the auxiliary space and in the \lq\lq quantum'' spaces we take
the spin-1 representation of SU(2).

It is a common experience in the theory of Bethe Ansatz equations that the 
polynomials $T_k$ have only real roots inside the physical strip 
$\vert{\rm Im}\, \theta\vert<q$ (see, for example, \cite{SuzukiS1}). 
We denote the set of real roots of $T_k$ (which may be empty) 
by $t_k$ for $k=1,2,\dots$. From (\ref{T0T1}) we see that
\begin{equation}
t_1=X\cup \tilde t_1,
\label{Xset}
\end{equation}
where $\tilde t_1$ is the set of real roots of $\tilde T_1$ and from
(\ref{yT}) we see that, denoting the set of roots of $y_k$ inside the
physical strip by $\eta_k$,
\begin{equation}
\eta_k=t_{k-1}\cup t_{k+1},\quad k=2,3,\dots\qquad \eta_1=t_2.
\label{etas}
\end{equation}
Furthermore, we see that the quantization conditions 
\begin{equation}
Y_k(z-iq)=0,\quad z\in t_k,\quad k=2,3,\dots\qquad
Y_1(z-iq)=0,\quad z\in \tilde t_1
\label{yQC}
\end{equation}
are satisfied by (\ref{YT}). 

We need one more equation to quantize the rapidities of the physical particles.
They are determined from $y_0(\theta)$ by the equation:
\begin{equation}
Y_0(\theta_k-iq)=0,\quad k=1,2,\dots N
\label{BYeqs}
\end{equation}
For large $\ell$ this equation is equivalent to the Bethe-Yang equations
\begin{equation}
{\rm e}^{i\ell{\rm sinh}\theta_k}\Lambda(\theta_k\vert\underline\theta)=-1,
\end{equation}
if $y_0(\theta)={\rm e}^{-\ell{\rm cosh}\theta}y_1(\theta)$.This relation is consistent
with the Y-system equations (\ref{Y}) independently of the value of $\ell$.
In this paper we will assume that this relation holds exactly.

For finite $\ell$ the Y-system equations (\ref{Ysys}) for $k=3,4,\dots$ 
and (\ref{Ysys1}) are unchanged and only the $k=2$ equation is modified to
\begin{equation}
y_2(\theta+iq)\,y_2(\theta-iq)=Y_3(\theta)\,Y_1(\theta)\,Y_0(\theta).
\label{Ysys2}
\end{equation}
Since the only change is the inclusion of the multiplyer $Y_0$, which is
very close to unity for large $\ell$, we will
assume that the finite $\ell$  solution is a smooth deformation
of the Bethe Ansatz construction, i.e. both the functions $y_k(\theta)$
and the sets $t_k$ are smooth deformations of the ones determined by the
above explicit construction at large $\ell$. (\ref{Xset}) still holds,
but with $X$ containing the modified particle rapidities $\theta_k$.

Now it would be possible to rewrite the Y-system equations for excited states
as a set of TBA integral equations similar to (\ref{TBA}), by adding source
terms corresponding to the $y_k$ roots. This would be supplemented by
quantization conditions of the form (\ref{yQC}) and (\ref{BYeqs}). In this
paper we leave out this step and proceed directly to the NLIE description.

Assuming that the solution of the excited state Y-system (\ref{Y}) is
found, we first build a T-system, which is equivalent to it. Since the 
semi-infinte Y-system (\ref{Ysys}) for $k=3,4,\dots$ is of the form of the 
standard $A_1$ Y-system, we can find suitable T-system functions $\phi,T_k$ 
satisfying (\ref{T}) so that
(\ref{yT}) and (\ref{YT}) hold for $k=2,3,\dots$. The $k=2$ relation
(\ref{Ysys2}) corresponds to the modified relations
\begin{equation} \label{modyT}
Y_1(\theta)\, Y_0(\theta)-1=y_1(\theta)\,\left[Y_0(\theta)+{\rm e}^{-\ell{\rm cosh}\theta}\right]
=\frac{T_0(\theta) \, T_2(\theta)}{\phi(\theta+i 
q) \, \bar{\phi}(x-iq)},
\end{equation}
\begin{equation} \label{modYT}
Y_1(\theta)\, Y_0(\theta)=\frac{T_1(\theta+ i \, q) \, T_1(\theta-i \, q)}
{\phi(\theta+iq) \, \bar{\phi}(x-iq)}.  
\end{equation}
In Appendix A it is shown explicitly that it is always possible to find
T-system functions $T_k$ for $k=0,1,\dots$ such that (\ref{modyT}),
(\ref{modYT}) are satisfied, (\ref{yT}) and (\ref{YT}) hold for $k=2,3\dots$ 
and the roots of $T_k$ coincide with the elements of $t_k$ for $k=1,2,\dots$
Furthermore the T-system equations (\ref{T}) are satisfied for $k=1,2,\dots$
with $\phi$ given by (\ref{phi}) (with the modified $\theta_k$ rapidities).
The T-system functions obtained this way are smooth deformations of the
corresponding ones constructed in the large $\ell$ limit.

Next we construct the TQ system. This is based on the fact that
the T-system equations (\ref{T}) are integrable and have a Lax representation
\cite{KaziO4} through the auxiliary problem 
(TQ-relations)
\begin{eqnarray}
\!\!\!\!\!\!\!\!T_{k+1}(\theta)  Q(\theta \! + \! i k  q) \! - 
\!T_{k}(\theta \!-\!i q) 
Q(\theta\!+\!i (k\!+\!2) q)\!\!\!\!&=&\!\!\!\!
\phi(\theta\!+\!i  k q) \bar{Q}(\theta\!-\!i (k\!+\!2) q), \nonumber \\
\!\!\!\!\!\!\!\!T_{k-1}(\theta) \bar{Q}(\theta \!-\!i  (k\!+\!2) q)-T_k(\theta\!-\!i  q)  
\bar{Q}(\theta\!-\!i k q) 
\!\!\!\!&=&\!\!\!\!
-\bar{\phi}(\theta\!-\!i  k q) Q(\theta\!+\!i k q). \label{TQ}
\end{eqnarray}

It can be shown that in the Bethe Ansatz limit $Q(\theta)=\bar Q(\theta)=
Q_0(\theta)$ satisfies the TQ system (\ref{TQ}).
In Appendix A we show how to find a $Q(\theta)$ solution which is a smooth
deformation of $Q_0(\theta)$.

Both equations (\ref{T}) and (\ref{TQ}) are invariant under the gauge 
transformation
\begin{eqnarray}
T_{k}(\theta) &\rightarrow& g(\theta+i \, k q)\, \bar{g}(\theta-i \, k \, q) 
\, T_{k}(\theta), \nonumber \\
\phi(\theta) &\rightarrow& g(\theta+i \, q) \, g(\theta-i \, q) 
\, \phi(\theta), \nonumber \\
Q(\theta) &\rightarrow& g(\theta-i \, q) \, Q(\theta). \label{gauge}
\end{eqnarray}
The Y-functions $y_k$ are gauge invariant.

We will use for the NLIE the following gauge invariant auxiliary functions:
\begin{eqnarray}
b_k(\theta) &=& \frac{Q(\theta+i \, (k+2) 
\, q)}{\bar{Q}(\theta-i \,(k+2) \, q)}
\frac{T_k(\theta-i \, q)}{\phi(\theta+i \, k \, q)}, \qquad k \geq 0, 
\label{bk}  \\
B_k(\theta)&=&\frac{Q(\theta+i \, k \, q)}{\bar{Q}(\theta-i \, (k+2)\,q)}
\frac{T_{k+1}(\theta)}{\phi(\theta+i \, k \, q)} \qquad k \geq 0,  \label{Bk}
\end{eqnarray}
where 
\begin{equation}
B_k(\theta)=1+b_k(\theta). 
\label{bB}
\end{equation}
Furthermore the auxiliary functions (\ref{bk},\ref{Bk}) 
satisfy the functional equations
\begin{eqnarray}
b_k(\theta) \, \bar{b}_k(\theta)&=&Y_k(\theta)
\,[1+\delta_{k1}y_0(\theta)], \qquad k \geq 1, \label{bbbar}
 \\
B_k(\theta+i \, q) \, \bar{B}_k(\theta-i \,q)&=&Y_{k+1}(\theta),
 \qquad k \geq 1. \label{BBbar}
\end{eqnarray} 
(\ref{bB}), (\ref{bbbar}) and (\ref{BBbar}), together with the analytic
properties that can be read off the reperesentation (\ref{bk},\ref{Bk})
are the key to find a set of NLIEs that effectively
truncate the TBA equations at the $k$th node.

The complete set of unknown functions of our NLIE we discuss below
are the functions: $b_1(\theta)$, which will be 
denoted by $b(\theta-i \, \gamma)$ for short, 
$y_1(\theta)$ and $y_0(\theta)$. For the quantization conditions 
the function $b_0(\theta)$ will be used as well.

The derivation of the NLIE is given in Appendix B, for an important
subset of multiparticle states.
Here we  present the equations and quantization
conditions for the most general excited state of the model precisely.
The NLIEs take the form:
\begin{eqnarray}
\log b(\theta) &=& \! \! i \, \pi \, \delta_b+i D^{+ \gamma} (\theta)+i g_1^{+ \gamma}(\theta)
+i g_b^{+ \gamma}(\theta)+(G* \ln B)(\theta)-
 (G^{+2 \gamma}* \ln {\bar B})(\theta) \nonumber \\
&+& (K^{-\frac{\pi}{2}+\gamma}*\ln Y_1)(\theta)+(K^{-\frac{\pi}{2}+\gamma}*\ln Y_0)(\theta), 
\qquad \qquad \qquad  \qquad \delta_b \in \{0,1\}  \nonumber \\
\log y_1(\theta)&=&\! \! i \, \pi \delta_y+i g_y(\theta)+(K^{+\frac{\pi}{2}-\gamma}* \ln B)(\theta)+
 (K^{-\frac{\pi}{2}+\gamma}* \ln {\bar B})(\theta), \qquad \delta_y \in \{0,1\} \nonumber \\
\log y_0(\theta)&=& -\ell \, \cosh \theta + \log y_1(\theta), \nonumber \\
B(\theta) &=& 1+b(\theta), \quad Y_1(\theta)=1+y_1(\theta), \quad Y_0(\theta)=1+y_0(\theta). \label{nlie1}
\end{eqnarray}
where $\gamma$ is a contour shifting parameter restricted into the interval $0<\gamma<\frac{\pi}{2},$
\newline $\ln$ denotes the "fundamental" logarithm function having its branch cut on the negative real axis
and we introduced the notation for any function $f$:
$$ f^{\pm \eta}(\theta)=f(\theta\pm i \eta).$$
The kernel function $G$  of (\ref{nlie1}) reads as
  \begin{equation} \label{GK}
 G(\theta)=\int\limits _{-\infty}^{\infty}\frac{{\rm d}q}{2\pi}\,\,
e^{iq\theta}\,\,\frac{e^\frac{-\pi |q|}{2}}{2\cosh\frac{\pi q}{2}},
\end{equation}
while $K(\theta)$ is the kernel of the TBA equations.
The source terms of (\ref{nlie1}) read as:
\begin{eqnarray}
g_{b}(\theta) & = & \sum_{j=1}^{N_{2}}\chi(\theta-h_{j})+
\sum_{j=1}^{N_{V}^{S}}\left(\chi(\theta-v_{j})+\chi(\theta-\bar{v}_{j})\right)
-\sum_{j=1}^{N_{S}}\left(\chi(\theta-s_{j})+\chi(\theta-\bar{s}_{j})\right) \nonumber \\
 & - & \sum_{j=1}^{M_{C}}\chi(\theta-c_{j})-\sum_{j=1}^{M_{W}}\chi_{II}(\theta-w_{j}),\\
g_{1}(\theta) & = & \sum_{j=1}^{N_{1}}\chi_{K}(\theta-h_{j}^{(1)}),\\
g_{y}(\theta) & = & \lim_{\eta\rightarrow 0^{+}}\tilde{g}_{y}\left(\theta+i\frac{\pi}{2}-i\eta\right),
\nonumber \\
\tilde{g}_{y}(\theta) & = & 
\sum_{j=1}^{N_2}\chi_{K}(\theta-h_{j})
+\sum_{j=1}^{N_{V}^{S}}\left(\chi_{K}(\theta-v_{j})+\chi_{K}(\theta-\bar{v}_{j})\right)
-\sum_{j=1}^{M_{S}}\left(\chi_{K}(\theta-s_{j})+\chi_{K}(\theta-\bar{s}_{j})\right) \nonumber \\
 & - & \sum_{j=1}^{M_{C}}\chi_{K}(\theta-c_{j})
 -\sum_{j=1}^{M_{W}}\chi_{KII}(\theta-w_{j}),
 \end{eqnarray}
where $\chi(\theta)$  and $\chi_K(\theta)$ are proportional to the odd primitives of the kernel functions  
 \begin{equation} \label{chi}
\chi(\theta)=2\pi\int\limits _{0}^{\theta}
{\rm d}x\,\, G(x),\qquad\chi_{K}(\theta)=2\pi\int\limits 
_{0}^{\theta}
{\rm d}x\,\, K(x), 
\end{equation}
 and the second determination of any function $f_{II}(\theta)$ is defined by:
\begin{equation}\label{detII}
f_{II}(\theta)= 
f(\theta)+f(\theta-i\, \pi\,\mbox{sign}(\mbox{Im} \, \theta)). 
\end{equation}

The function $\chi_K(\theta)$ is given by the formula:
 \begin{equation} \label{}
\chi_K(\theta)=i \, \ln \frac{\sinh\left( i \frac{\pi}{4}+\frac{\theta}{2} \right)}
{\sinh\left( i \frac{\pi}{4}-\frac{\theta}{2} \right)}, \quad
\chi_K(\theta)=-\chi_K(-\theta) \quad  \forall \theta \in {\mathbb C}.
\end{equation}
The branch cuts are chosen to run parallel to the real axis so that $\chi_K(\theta)$ is an odd real 
analytic function on the entire complex 
plane and continuous along the real axis. In this case $\chi_K(\theta)$ is not periodic  anymore with 
respect to $2 \pi i$. It is periodic only modulo $2 \pi$, i.e. the following identity holds:
$$\chi_K(\theta+2\pi i)=\chi_K(\theta)-2\pi.$$ It follows that the distance between the consecutive cuts
is $2 \pi i$ and the jump of $\chi_K(\theta)$ is equal to $-2\pi$ at each branch cut crossed from below. 
The choice of branch cuts is depicted in Figure 2. 
\begin{figure}[htb]
\begin{flushleft}
\hskip 15mm
\leavevmode
\epsfxsize=120mm
\epsfbox{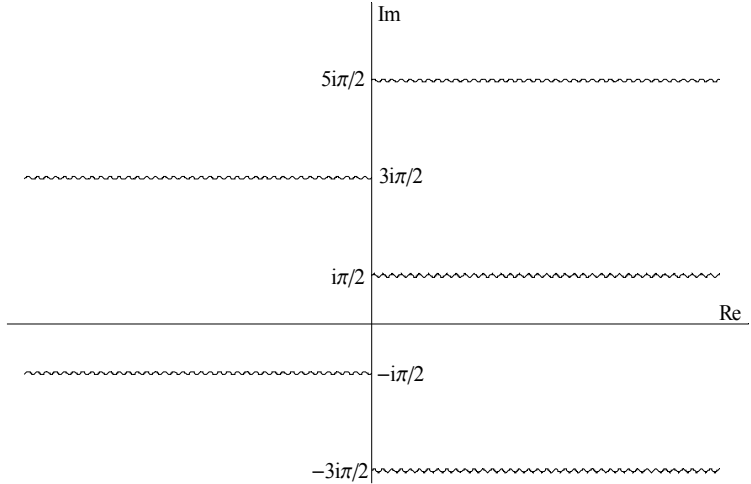}
\end{flushleft}
\caption{{\footnotesize
Locations of the branch cuts of $\chi_K(\theta).$
}}
\label{6}
\end{figure}

Although all physical quantities should be gauge-invariant, i.e. invariant
under (\ref{gauge}), the description of the source objects is nevertheless
more transparent in a particular gauge fixed by the condition (\ref{phi}).
In this gauge it is possible to show (see Appendix A) that $T_1(\theta_j)=0$ 
for all physical rapidity 
$\theta_j$,~$j=1,2,\dots,N$ and we can define
$\tilde T_1(\theta)$ by
\begin{equation}
T_1(\theta)=\tilde T_1(\theta)\varphi_0(\theta).
\end{equation}
In what follows we will work in this fixed gauge.

The objects appearing in the source terms of (\ref{nlie1}) are as follows.
\newline
1. \emph{Type I holes}\footnote{ The term "type I hole" comes from the fact that in the thermodynamic limit
of the equations
the real zeroes of $\tilde T_1(\theta)$ 
correspond to holes in the distribution of real Bethe roots.}
: \, $\{h_j^{(1)}\}, \quad j=1,\dots,N_1$, which are 
zeroes of $\tilde T_1(x)$
satisfying the condition: $0 \leq |\mbox{Im} \, h_j^{(1)}| < \frac{\pi}{2}$.
\newline
\newline
2. \emph{Holes}\footnote{ The term "hole" comes from the 
fact that in the thermodynamic limit
of the equations
the real zeroes of $T_2(\theta)$ correspond to holes in the distribution of 2-strings.}: \, $\{h_j\}, \quad 
j=1,\dots,N_2$ corresponding to zeroes of $T_2(x),$ with
$0 \leq |\mbox{Im} \, h_j| < \gamma$ 
\newline
\newline
3. "\emph{Close objects}": $\{c_j\}$ $\quad j=1,\dots,M_C$, which are "source objects"
satisfying the condition: $\gamma<|\mbox{Im} \, c_j|< \pi.$
\newline
\newline
4. "\emph{Wide objects}": \, $\{w_j\}, \quad j=1,\dots,M_W$, with $\pi<|\mbox{Im} \, w_j|.$
\newline
\newline
Close and wide objects appear in complex conjugate pairs and they are related to the zeroes of the $Q$ function 
in the
same manner as in the spin-1 XXX chains \cite{SuzukiS1,HegS1}.

There are also two types of special objects in the source terms of the equations. They are
defined by the relations:
\begin{equation} \label{sj}
\mbox{Im} \log b^{-\gamma}(s_j)=2 \pi I_{s_j}, \quad |b^{-\gamma}(s_j)|>1, \quad 
(\mbox{Im} \log b^{-\gamma})'(s_j)<0,
\quad j=1,\dots,N_S,
\end{equation}
\begin{equation} \label{vj}
\mbox{Im} \log b^{-\gamma}(v_j)=2 \pi I_{v_j}, \quad |b^{-\gamma}(v_j)|>1, \quad 
(\mbox{Im} \log b^{-\gamma})'(v_j)>0,
\quad j=1,\dots,N_V^S
\end{equation}
and they are called ordinary and virtual special objects respectively \cite{HegS1}.
 The function $D(\theta)$ of (\ref{nlie1}) is expressed by the
asymptotic rapidities of the $N$ particle state:
 $$D(\theta)=\sum\limits_{k=1}^N \chi_K(\theta-\theta_k).$$
The values of the constants of the NLIE (\ref{nlie1}) depend on the 
number of different source objects: 
\begin{equation}
 \delta_b\equiv\frac{N+N_1}{2}+J \,\, \mbox{(mod} \,\, 2), \qquad 
\mbox{and} \quad \delta_y\equiv J+M_W \,\, \mbox{(mod} \,\, 2),
\end{equation} 
where $J$ denotes the isospin of the state.
In addition to (\ref{nlie1}) we need two other equations for the determination of type I holes and 
wide objects.
For the determination of type I holes, a counting function can be defined: 
\begin{equation} \label{Z1}
Z_1(\theta)= \tilde g_y(\theta)+\frac{1}{i}(K^{-\gamma}* \ln B)(\theta)-
 \frac{1}{i}(K^{+\gamma}* \ln {\bar B})(\theta)-\delta_y, \qquad 0\leq |\mbox{Im} \,
 \theta|< \frac{\pi}{2},
\end{equation} 
where $Z_1(\theta)=-i \, \log b_0(\theta)$.
The function necessary for the determination of wide objects 
is
\begin{eqnarray} 
\log  b_0(\theta+iq) &=& i \, g_{bII}(\theta)+(G_{II}^{-\gamma}* \ln B)(\theta)
-(G_{II}^{+\gamma}* \ln {\bar B})(\theta), \quad \pi<\mbox{Im} \, \theta. \label{b1W}
\end{eqnarray}
Taking into account the fact that $K_{II}(\theta)$ and $\chi_{KII}(\theta)$ 
are zero, this equation looks as if it were the second determination 
of the first equation in (\ref{nlie1}).

The source objects appearing in the NLIE are not arbitrary parameters, 
they have to satisfy
certain quantization conditions dictated by the analytic properties of the unknown 
functions of the NLIE. The quantization conditions for "magnonic" degrees of freedom read as:
\begin{itemize}
\item For holes: \begin{equation}
\frac{1}{i}\,\log b^{-\gamma}(h_{j})=2\pi\, I_{h_{j}}, \qquad j=1,...,N_2.\label{eq:holes}\end{equation}
 
\item For close source objects (only for the upper part of the close pair): \begin{equation}
\frac{1}{i}\,\log b^{-\gamma}(c_{j}^{\uparrow})=2\pi\, I_{c_{j}^{\uparrow}}, \qquad 
j=1,...,M_{C}/2.\label{eq:close}\end{equation}
 
\item For wide objects (upper part of the wide pair): \begin{equation}
-i\,\log b_0(w_{j}^{\uparrow}+iq)=2\pi\, I_{w_{j}^{\uparrow}}, \qquad
j=1,...,M_{W}/2.\label{eq:wide}\end{equation}
  
\end{itemize}
(We have determined only the upper parts of the complex pairs,
the lower parts are simply given by complex conjugation.) 

For ordinary and virtual special objects the defining relations
(\ref{sj}) and (\ref{vj}) themselves play the role of
quantization conditions.

\begin{itemize}
\item Finally for type I holes: \begin{equation}
Z_1 (h_{j}^{(1)})=2\pi\, I_{h_{j}^{(1)}}, \qquad j=1,...,N_{1}.\label{eq:t1}\end{equation}

\end{itemize}

In finite volume the rapidities of physical particles are also quantized. 
Their quantized values can be determined from $y_0(\theta)$:
\begin{equation} 
\frac{1}{i} \log y_0(\theta_k + i \frac{\pi}{2})=2 \pi \, I_k,\qquad k=1,...,N \label{qc:theta_k}.
\end{equation}

All the above quantum numbers $I_{\alpha_{j}}$'s are half integers.
A state is then identified by a choice of the quantum numbers $(I_{h_{j}},I_{c_{j}},...)$.
We note that the NLIE itself can impose constraints on the allowed values 
of the magnonic quantum numbers. Moreover there are relations among the 
numbers of different species of source objects.
They are called counting equations and in this model there are two of them:
\begin{equation} \label{cet2}
N_2+2N_V^S-2N_S =M_C+2J+2 M_W,
\end{equation}
\begin{equation} \label{cet1}
N_1-2N_R^S=J+M_1-M_R,
\end{equation}
where $M_1$ denotes the number of close and wide objects lying farther  
than $\frac{\pi}{2}$ from the real axis, 
$M_R$ stands for the number of real\footnote{{Real roots are those 
zeroes of $1+b_0(\theta)$ along the real axis
which are not type I holes. They correspond to real solutions 
of the asymptotic Bethe equations.}}
 Bethe roots of the asymptotic Bethe Ansatz equations of 
the model, and $N_R^S$ means the number of the \emph{real special objects}.
Real special objects can be either real Bethe roots or real type I holes. They are called specials because
at the positions of real specials the counting function of real Bethe roots is no more monotonically  
increasing: i.e. $ \frac{d}{dx} Z_1 (x)<0$.
We note that the counting equation (\ref{cet1}) is valid when all the type I holes are real, while 
(\ref{cet2}) is valid also for complex holes as far as their imaginary parts lie within the
interval $[-\gamma,\gamma]$

In the TBA analysis of the sine-Gordon theory \cite{BH-SG} it turned out 
that the energy expression for an $N$-particle state 
consists of two terms: the sum of the kinetic
energies of $N$ non-interacting particles plus an integral 
expression similar to (\ref{TBAenergy}) containing the Y-function 
corresponding to the massive node of the Y-system.
Assuming this formula to be valid also for the $O(3)$ NLS model
the energy can be expressed in terms of the solutions of the NLIE:
\begin{equation} \label{energy}
E= \sum\limits_{k=1}^N \cosh \theta_k -\int\limits_{-\infty}^{\infty} 
\frac{{\rm d}\theta}{2 \pi}
\, \cosh \theta \, \ln Y_0(\theta).
\end{equation}

\section{NLIE in the IR limit and solution of the asymptotic Bethe equations
for 2-particle states }

In the $\ell \to \infty$ limit the NLIE for $N$ 
particles reduces to the NLIE of an 
inhomogeneous N-site spin-1 $SU(2)$ vertex model \cite{SuzukiS1,HegS1}, 
where the roles of inhomogeneities are 
played by the physical rapidities. The NLIE description for the inhomogeneous 
spin-1 $SU(2)$ vertex model 
is equivalent to the Bethe Ansatz solution, thus in the infrared (IR) limit 
the NLIE (\ref{nlie1}) is 
equivalent to the asymptotic Bethe equations~(\ref{uj}).
The relation between the magnon rapidities in (\ref{uj}) 
and the source objects of the NLIE is as follows. The 
magnons with $|\mbox{Im} \, u_j|>\frac{\pi}{2}+\gamma$ can be 
obtained by the formula $u_j=U_j+i \, 
\frac{\pi}{2} \, \mbox{sign}(\mbox{Im}\, U_j)$, where
$U_j$ is the infrared limit of a close or wide object. 
Magnons with $|\mbox{Im} \, u_j|<\frac{\pi}{2}+\gamma$
are summed up by the integral terms of the NLIE this is why no 
source terms can be associated to them. 

In this limit the quantization condition (\ref{qc:theta_k}) 
is equivalent to the asymptotic
Bethe-Yang quantization for the physical rapidities:
\begin{equation} \label{BY}
{\rm e}^{i \ell \, \sinh \theta_k} \, 
\frac{\varphi_0(\theta_k-i \, \pi)}{\varphi_0(\theta_k+i \, \pi)}
\frac{Q_0(\theta_k+i \, \pi)}{Q_0(\theta_k-i \, \pi)}=-1 \qquad k=1,..,N.
\end{equation} 
The term multiplying ${\rm e}^{i \ell \, \sinh \theta_k}$ on the left hand 
side of (\ref{BY}) is nothing but
the eigenvalue of the "color" transfer matrix the $O(3)$ NLS model taken at the
point $\theta_k$:
\begin{equation} \label{lambdaO3}
\frac{\varphi_0(\theta_k-i \, \pi)}{\varphi_0(\theta_k+i \, \pi)}
\frac{Q_0(\theta_k+i \, \pi)}{Q_0(\theta_k-i \, \pi)}=
\Lambda(\theta_k,\underline\theta).
\end{equation}
Thus the rapidity quantization conditions given by the NLIE 
(\ref{nlie1}-\ref{qc:theta_k}) 
in the IR limit are identical to the Bethe-Yang equations of the 
$O(3)$ NLS model. 

\subsection*{2-particle states}

In this subsection we analyse the structure of the asymptotic solution
of the NLIE equations based on the explicit T-system solution (\ref{asyT}).
We consider zero momentum 2-particle states, with states corresponding to
$J=0,1,2$ in the isospin space. The analytic properties of the
problem are determined by the magnonic Bethe roots
and the zeroes of the transfer matrices $\tilde T_1(\theta)$ and $T_2(\theta)$ 
inside the physical strip{\footnote {The strip $|\mbox{Im}\theta|\leq 
\frac{\pi}{2}$ in the complex plane.}}.  
>From (\ref{asyT}) we see that the above data depend on $J$ and the two
physical rapidities $\theta_{1,2}$ and for zero-momentum (symmetric)
states with fixed $J$ the state can be characterized by the 
magnitude of $\theta_1$.

The result of the calculation is as follows.
\newline

{\bf \underline{$J=2$ state:}} \newline
For all real values of the rapidities $\theta_1$ and $\theta_2$:
\begin{itemize}

\item $Q_0(\theta)=1,$

\item $\tilde T_1(\theta)$ has two real zeroes,

\item $T_2(\theta)$ has four real zeroes.

\end{itemize}

{\bf \underline{$J=1$ state:}} \newline
For any symmetric state $\theta_1=-\theta_2$: $Q_0(\theta)=\theta,$ and 
the distribution of zeroes of the
transfer matrices depend on the magnitude of $\theta_1$. \newline

 $0<\theta_1<\frac{\sqrt{3}}{2}\pi\simeq2.72070$ 
\begin{itemize}

\item $\tilde T_1(\theta)$ has no zeroes,

\item $T_2(\theta)$ has two real zeroes.

\end{itemize}

$\frac{\sqrt{3}}{2}\pi<\theta_1<\pi\simeq3.14159$ 
\begin{itemize}

\item $\tilde T_1(\theta)$ has two complex zeroes,

\item $T_2(\theta)$ has two real zeroes.

\end{itemize}

$\pi<\theta_1<\sqrt{\frac32 +\frac12 \sqrt{\frac{43}{3}}} \pi \simeq 5.78682$ 
\begin{itemize}

\item $\tilde T_1(\theta)$ has two real zeroes,

\item $T_2(\theta)$ has two real zeroes.

\end{itemize}

$\sqrt{\frac32 +\frac12 \sqrt{\frac{43}{3}}} \pi<\theta_1< \frac{\sqrt{15}}{2} 
\pi \simeq 6.08367$ 

\begin{itemize}

\item $\tilde T_1(\theta)$ has two real zeroes,

\item $T_2(\theta)$ has two real and two complex zeroes.

\end{itemize}

$\frac{\sqrt{15}}{2} \pi<\theta_1$ 

\begin{itemize}

\item $\tilde T_1(\theta)$ has two real zeroes,

\item $T_2(\theta)$ has four real zeroes and there is a special object 
in the corresponding NLIE as well.

\end{itemize}

{\bf \underline{$J=0$ state:}} \newline
For this state the $Q_0$ function in the IR limit reads as:
\begin{equation} \label{QJ0}
Q_0(\theta)=(\theta-u_0+i \zeta)(\theta-u_0-i \zeta), 
\qquad u_0=\frac{\theta_1+\theta_2}{2}, 
\end{equation}
where $
\zeta=\frac12 \sqrt{\frac{4 \pi^2}{3}+\frac{(\theta_1-\theta_2)^2}{3}}. $
Depending on the magnitude of the rapidities for this state 
there are seven regions corresponding
to the various possibilities for the  distribution of the relevant 
zeroes of the transfer matrices $\tilde T_1$ and $T_2$.
In our numerical investigations for $J=0$ we considered only the simplest 
symmetric states with 
$0<\theta_1<\sqrt{\frac{17-\sqrt{241}}{6}} \pi \simeq 1.55809$. 
In this case neither $\tilde T_1$ nor
$T_2$ has zeroes in the physical strip.

\section{Numerical results}

We have considered the NLIE description of the cases $N=0$ (vacuum) and 
$N=1$ (mass gap) previously \cite{BH34,HegO34}. Some numerical results 
for these cases are given in the $\varepsilon_0,\varepsilon_1$ columns of
Table \ref{tabX}.

We now discuss the $N=2$ (2-particle) cases extensively. 
This is the next non-trivial case, with possible isospin values $J=0,1,2$.  
Concretely, we consider only zero momentum states with physical
rapidities $\theta_{1,2}=\pm H$ with the smallest possible momentum
values. In addition, we also consider some special $N=3$, $J=1$
3-particle states with physical rapidities $\theta_{1,3}=\pm H$,
$\theta_2=0$, which leads to an NLIE very similar to the one corresponding
to the RI1b region for 2-particle states discussed below.

In all cases we started from the large $\ell$ BY solution and found the
numerical solution of the NLIE equation by iteration. We used the following
parameters:

$h=0.04$ (step size for the Simpson formula in rapidity space)

$h^\prime=0.004$ (step size for the Fourier integrals)

$\Lambda=240$ (cutoff in rapidity space)

$\epsilon=10^{-12}$ ((absolute) accuracy for solving the quantization
conditions)

With these parameters, it was necessary to perform a few hundred iterations
(corresponding to less than 1 hour CPU on a laptop). In this way a (relative)
accuracy of about $2\cdot 10^{-8}$ is achieved for both the energy values and
the position of the roots, except very close to the boundary points between
regions where qualitative changes occur (discussed below).

For $N=2$, $J=2$ there is just one region: starting from large $\ell$, we 
can gradually come down to the UV region $\ell\sim10^{-6}$ without any 
qualitative change in the algorithm. The qualitative description is correctly
given by the BY-equations: there are two pairs of real roots for $T_2$
(arranged symmetrically around the origin) and one pair of real roots for
$\tilde T_1$ (also symmetrical).
Some numerical results are given in Table \ref{tabX} ($\varepsilon_{22}$
column).

\begin{table}[h]
\centering 
\begin{tabular}[h]{c|c|c|c|c} 
$\ell$   &  $\varepsilon_0$  &  $\varepsilon_1$   &  $\varepsilon_{22}$ &
$\varepsilon_{21}$  \\
\hline \hline 
0.000001 & -0.96553941 & -0.64792555 & -0.01287029 & 11.59612929 \\
\hline
0.000003 & -0.96027050 & -0.62282893 &  0.05183255 & 11.60073719 \\
\hline
0.00001 & -0.95362228 & -0.59128817 &  0.13308151 & 11.60648275 \\
\hline
0.00003 & -0.94655984 & -0.55793920 &  0.21890220 & 11.61250075 \\
\hline
0.0001 & -0.93741883 & -0.51502162 &  0.32920544 & 11.62015568 \\
\hline
0.0003 &  -0.92740876 & -0.46835316 &  0.44895233 & 11.62835934 \\
\hline
0.001 & -0.91395474 & -0.40619591 &  0.60808718 & 11.63907639 \\ 
\hline
0.003 & -0.89850538 & -0.33567869 &  0.78805971 & 11.65090156 \\
\hline
0.01 &  -0.87635700 & -0.23640710 &  1.04016160 & 11.66669458 \\
\hline
0.03 & -0.84838500 & -0.11495754 &  1.34580545 & 11.68338729 \\
\hline
0.1 & -0.80069857 &  0.07772108 &  1.82023294 & 11.69620904 \\
\hline
1.0 &  -0.48624957 &  1.08420867 &  3.96980668 & 11.60661538 \\
\hline
\end{tabular} 
\caption{\footnotesize Finite size spectrum of the O$(3)$ model.
The $\varepsilon_0=\ell\,E_0$ column corresponds to the vacuum state,
the $\varepsilon_1=\ell\,E_1$ column to the (standing) one-particle
state, and the two last columns correspond to the zero-momentum
$N=2,\ J=2$ and $N=2,\ J=1$ states, respectively. 
}
\label{tabX} 
\end{table}

The case $N=2$, $J=1$ is more complex. From the large $\ell$ BY equations we
see that in this case only $T_2$ has a single pair of real roots. Let us 
denote this region RI1a. From the BY equations we get the estimate
$\ell\gtrsim0.74$, below which $\tilde T_1$ has (a complex conjugate pair of)
imaginary roots inside the physical strip. Indeed, we can monitor the 
imaginary $\tilde T_1$ roots as they move towards the physical strip by solving
the RI1a NLIE equations for decreasing $\ell$ values and we estimate that the
boundary of RI1a is at $\ell\cong0.65$, below which (region RI1c) we have 
to solve the NLIE with one complex conjugate pair of imaginary roots for 
$\tilde T_1$ and one pair of real roots for $T_2$. 
The limits of this region are
\begin{equation}
0.37\lesssim\ell\lesssim0.65
\end{equation}
(as opposed to the $0.49\lesssim\ell\lesssim0.74$ estimate from the BY
solution).

At $\ell=0.37$ the imaginary $\tilde T_1$ roots meet 
at the origin and as $\ell$
decreases further they move away symmetrically from the origin along the
real axis. We denote this region by RI1b. According to the BY solution, RI1b 
ends at $\ell\cong0.036$ but the BY equations are no longer relevant for 
such small $\ell$ values and in fact we find that RI1b extends all the way 
to the UV limit $\ell\to0$. The three regions of the $N=2$, $J=1$ problem 
are summarized in Table \ref{tabR}.
Some numerical results are given in Table \ref{tabX} ($\varepsilon_{21}$
column).

\begin{table}[h]
\centering 
\begin{tabular}[h]{c|c|c|c} 
region   &  range  &  $\tilde T_1$ roots  &  $T_2$ roots \\
\hline \hline 
RI1a & $\ell>0.65$ & -- & 1 real pair \\
\hline
RI1c & $0.37<\ell<0.65$ & 1 imaginary pair & 1 real pair \\
\hline
RI1b & $\ell<0.37$ & 1 real pair & 1 real pair \\
\hline
\end{tabular} 
\caption{\footnotesize The regions in the $N=2$, $J=1$ case.
}
\label{tabR} 
\end{table}

For $N=2$, $J=0$ we have not considered all regions. The large $\ell$
RI0a region is characterized by no roots for either $\tilde T_1$ or $T_2$. It
corresponds to $\ell\gtrsim2$, below which an imaginary pair of $T_2$
roots enters the physical strip. The $N=2$, $J=0$ problem corresponds
to a number of different regions, but this is not discussed here in detail.

\subsection*{Rotator spectrum}

At small physical volumes, $\ell\to0$, the only relevant degrees of freedom 
are those corresponding to the zero modes of the $\sigma$ fields and the 
low lying energy levels are the eigenvalues of the effective Hamiltonian
\cite{L}
\begin{equation}
{\cal H}_{\rm eff}=\frac{1}{2\Theta_{\rm eff}}\hat L^2,
\end{equation}
where $\hat L^2$ is the quadratic Casimir operator with eigenvalues
$J(J+1)$ and $\Theta_{\rm eff}$ is the effective moment of inertia (which 
depends on the volume $\ell$). More precisely, the lowest energy levels in 
the isospin $J$ sector are given as\footnote{Note that $\tilde E_J(\ell)=
E_J(\ell)-E_0(\ell)$, i.e. the energy eigenvalues are measured from the lowest energy state (vacuum).}
\begin{equation}
\tilde\varepsilon_J(\ell)=\ell\tilde E_J(\ell)\cong g(\ell)\,J(J+1),
\end{equation}
where $g(\ell)$ is some effective running coupling. This effective
rotator spectrum is valid in perturbation theory (PT) up to 2-loop level
\cite{NWS,FP}. The lowest energy levels correspond to the totally symmetric 
tensor states, which in our notation are the $J=N$ states (diagonal 
scattering or $U(1)$ sector).

The $J=1$ case corresponds to the mass gap. In this case the PT result is 
known up to 3-loop order \cite{Shin}:
\begin{equation}
\tilde\varepsilon_1(\ell)=2\pi\alpha(1+\alpha^2+1.19\,\alpha^3)+
O(\alpha^5),
\label{varep1}
\end{equation}
where $\alpha$ is a convenient running coupling defined by
\begin{equation}
\frac{1}{\alpha}+\ln\alpha=\ln\frac{32\pi}{\ell}+\Gamma^\prime(1)-1.
\end{equation}
Here the constants on the right hand side of this equation are chosen such 
that the 1-loop term in (\ref{varep1}) vanishes. For general isospin
no direct calculation exists beyond 1-loop order \cite{NWS}, but using the
validity of the rotator spectrum up to 2-loop order \cite{FP} we have
\begin{equation}
\tilde\varepsilon_J(\ell)=J(J+1)\pi\alpha(1+\alpha^2)+
O(\alpha^4).
\label{varepI}
\end{equation}
Using the numerical results in Table \ref{tabX}, we established that
\begin{equation}
\tilde\varepsilon_2(\ell)=6\pi\alpha(1+\alpha^2)+
O(\alpha^5),
\label{varep2}
\end{equation}
i.e. the 3-loop term in (\ref{varep2}) vanishes or is very small.

The excited states above the rotator spectrum start at
\begin{equation}
\tilde\varepsilon^*(\ell)\cong4\pi,
\end{equation}
corresponding to two massless particles of momentum $\pm\frac{2\pi}{\ell}$.

\begin{figure}
\begin{center}
\psfig{figure=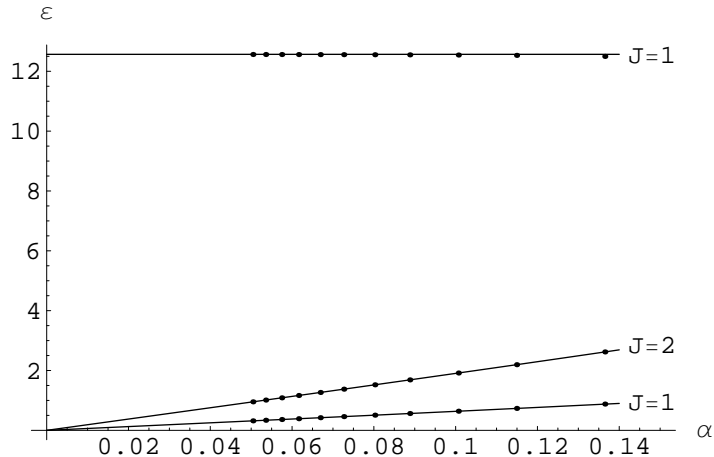,width=10cm}
\end{center}
\vspace{-0.5cm}
\caption{\footnotesize
The UV spectrum of the O$(3)$ model as function of the  running coupling 
$\alpha$, corresponding to the range $0.000001<\ell<0.1$. The two low 
energy data sets contain the $J=1$ and $J=2$ rotator energy levels 
and the corresponding solid lines are the PT predictions (\ref{varep1}) 
and (\ref{varep2}), respectively. Far above the rotator spectrum energies
of the $N=2$, $J=1$ excited states are shown. Here the solid line is the 
constant $4\pi$. 
}
\label{rotator}
\end{figure}

The UV spectrum of the $\sigma$ model, obtained by numerically solving the 
NLIE equations down to $\ell=10^{-6}$ agrees very well with the above 
rotator picture. This is shown in Figure \ref{rotator}.

\subsection*{The Bajnok-Janik formula}

Recently \cite{BJ} a generalization of L\"uscher's well-known result 
\cite{Luscher} about the finite size corrections of particle masses
has been suggested. According to this conjecture, in 2-dimensional 
relativistic integrable models\footnote{For the modifications necessary 
for the nonrelativistic model relevant in the AdS/CFT problem see \cite{BJ}.} 
one first defines the auxiliary function
\begin{equation}
{\cal F}(\underline\theta)=\frac{1}{2\pi}\int_{-\infty}^\infty\,
{\rm d}y\,{\rm e}^{-\ell\cosh y}\,\Lambda(y+iq\vert\underline\theta),
\label{BJ0}
\end{equation}
where $\underline\theta=(\theta_1,\theta_2,\dots,\theta_N)$ are the 
physical rapidities of an $N$-particle state and
$\Lambda(\theta\vert\underline\theta)$ is the eigenvalue of the
transfer matrix corresponding to the state in question.

Then the leading (for large $\ell$) finite size correction to the
particle rapidieties are given by the solution of the modified
Bethe-Yang equations
\begin{equation}
{\rm e}^{i\ell\sinh\theta_j}\,\Lambda(\theta_j\vert\underline\theta)\left\{
1+i\frac{\partial{\cal F}}{\partial\theta_j}\right\}=-1,
\qquad\quad j=1,2,\dots,N.
\label{BJ1}
\end{equation}
Finally the energy formula, which includes the leading FS corrections
is given by
\begin{equation}
E=\sum_{j=1}^N\,\cosh\theta_j-\frac{1}{2\pi}\int_{-\infty}^\infty\,
{\rm d}y\,\cosh y\,
{\rm e}^{-\ell\cosh y}\,\Lambda(y+iq\vert\underline\theta).
\label{BJ2}
\end{equation}
Here the particle rapidities $\theta_j$ are solutions of the modified
BY equations (\ref{BJ1}).

Since the above conjecture has been questioned for states with non-diagonal
scattering in the case of the O$(4)$ model \cite{KaziO4}, we decided to
numerically investigate the validity of it in our model. The results
are shown in Tables \ref{tab1}-\ref{tab4}. In all cases we studied (2-particle
states with isospin 0,1,2 and a 3-particle state with isospin~1) (\ref{BJ1})
and (\ref{BJ2}) seem to correctly give the leading FS corrections for
the particle rapidities and the energy of the state, respectively.

\begin{table}[h]
\centering 
\begin{tabular}[h]{c||c|c|c||c|c|c} 
$\ell$  &  $H^{(0)}$  &  $H^{(1)}$   &  $H$ &  $E^{(0)}$ & $E^{(1)}$ & $E$ \\
\hline \hline 
1 & 1.322083 & 1.316882 & 1.317113 & 4.017806 & 3.965831 & 3.969807 \\
\hline
2 & 0.912509 & 0.911777 & 0.911788 & 2.892078 & 2.881859 & 2.882081 \\
\hline
3 & 0.705671 & 0.705531 & 0.705532 & 2.518982 & 2.516186 & 2.516206 \\
\hline
4 & 0.576883 & 0.576852 & 0.576852 & 2.342126 & 2.341269 & 2.341271 \\
\hline
\end{tabular} 
\caption{\footnotesize 
Test of the Bajnok-Janik conjecture for zero momentum $N=2$, $J=2$
states. $H^{(0)}$ is the rapidity from the BY equation, $H^{(1)}$
from the modified BY equation (\ref{BJ1}) and $H$ is the exact rapidity.
The energy values $E^{(0)}$, $E^{(1)}$ and $E$ are defined analogously.
}
\label{tab1} 
\end{table}

\begin{table}[h]
\centering 
\begin{tabular}[h]{c||c|c|c||c|c|c} 
$\ell$  &  $H^{(0)}$  &  $H^{(1)}$   &  $H$ &  $E^{(0)}$ & $E^{(1)}$ & $E$ \\
\hline \hline 
1 & 2.424832 & 2.426304 & 2.426517 & 11.388820 & 11.579614 & 11.606615 \\
\hline
2 & 1.756669 & 1.757402 & 1.757428 & 5.965728 & 6.006502 & 6.008211 \\
\hline
3 & 1.392858 & 1.393083 & 1.393086 & 4.274704 & 4.285133 & 4.285278 \\
\hline
4 & 1.155252 & 1.155316 & 1.155317 & 3.489801 & 3.492757 & 3.492771 \\
\hline
\end{tabular} 
\caption{\footnotesize 
Test of the Bajnok-Janik conjecture for zero momentum $N=2$, $J=1$
states. $H^{(0)}$ is the rapidity from the BY equation, $H^{(1)}$
from the modified BY equation (\ref{BJ1}) and $H$ is the exact rapidity.
The energy values $E^{(0)}$, $E^{(1)}$ and $E$ are defined analogously.
}
\label{tab2} 
\end{table}

\begin{table}[h]
\centering 
\begin{tabular}[h]{c||c|c|c||c|c|c} 
$\ell$  &  $H^{(0)}$  &  $H^{(1)}$   &  $H$ &  $E^{(0)}$ & $E^{(1)}$ & $E$ \\
\hline \hline 
2 & 1.444416 & 1.448838 & 1.448743 & 4.475261 & 4.464046 & 4.465255 \\
\hline
3 & 1.055912 & 1.057006 & 1.056993 & 3.222472 & 3.212762 & 3.212965 \\
\hline
4 & 0.818360 & 0.818622 & 0.818621 & 2.707934 & 2.703854 & 2.703885 \\
\hline
5 & 0.661971 & 0.662035 & 0.662035 & 2.454443 & 2.452936 & 2.452940 \\
\hline
\end{tabular} 
\caption{\footnotesize 
Test of the Bajnok-Janik conjecture for zero momentum $N=2$, $J=0$
states. $H^{(0)}$ is the rapidity from the BY equation, $H^{(1)}$
from the modified BY equation (\ref{BJ1}) and $H$ is the exact rapidity.
The energy values $E^{(0)}$, $E^{(1)}$ and $E$ are defined analogously.
}
\label{tab3} 
\end{table}

\begin{table}[h]
\centering 
\begin{tabular}[h]{c||c|c|c||c|c|c} 
$\ell$  &  $H^{(0)}$  &  $H^{(1)}$   &  $H$ &  $E^{(0)}$ & $E^{(1)}$ & $E$ \\
\hline \hline 
1 & 2.441754 & 2.443044 & 2.443196 & 12.580190 & 12.762041 & 12.786246 \\
\hline
2 & 1.786837 & 1.787192 & 1.787205 & 7.138026 & 7.176748 & 7.178369 \\
\hline
3 & 1.425225 & 1.425320 & 1.425321 & 5.399247 & 5.409763 & 5.409915 \\
\hline
4 & 1.185639 & 1.185665 & 1.185665 & 4.578328 & 4.581482 & 4.581498 \\
\hline
\end{tabular} 
\caption{\footnotesize 
Test of the Bajnok-Janik conjecture for zero momentum $N=3$, $J=1$
states. $H^{(0)}$ is the rapidity from the BY equation, $H^{(1)}$
from the modified BY equation (\ref{BJ1}) and $H$ is the exact rapidity.
The energy values $E^{(0)}$, $E^{(1)}$ and $E$ are defined analogously.
}
\label{tab4} 
\end{table}

 \vspace{1cm}
{\tt Acknowledgements}

\noindent 
This investigation was supported by the Hungarian National Science Fund 
OTKA (under T049495).

\newpage

\appendix
\section{Construction of the T-system and TQ-system}

In this appendix we construct T-system elements satisfying (\ref{phi}),
(\ref{T}) for $k\geq1$, (\ref{yT}) and (\ref{YT}) for $k\geq2$ and the
modified relations (\ref{modyT}) and (\ref{modYT}), assuming that a
solution of the O$(3)$ Y-system is found satisfying (\ref{Y}) and the
Y-functions have roots corresponding to the sets (\ref{etas}).

We start by constructing the functions $\hat T_2$ and $\hat T_3$
satisfying
\begin{equation}
\hat T_k(\theta+iq)\,\hat T_k(\theta-iq)=Y_k(\theta)
\label{fund}
\end{equation}
for $k=2$, $k=3$ with the additional assumption that they are bounded for
large $\theta$ and their roots are the elements of the sets $t_2$, $t_3$.
The solution of this problem is the fundamental problem  in the theory of 
the TBA integral equations and is given explicitly by ($k=2,3$)
\begin{equation}
\hat T_k(\theta)=\prod_{\alpha}\,{\rm tanh}\,
\left(\frac{\theta-t_k^{(\alpha)}}{2}\right)\,{\rm exp}\left\{
\frac{1}{2\pi}\,\int_{-\infty}^\infty\frac{{\rm d}u}{{\rm cosh}(\theta-u)}\,
{\rm ln}\,Y_k(u)\right\},
\end{equation}
where $t_k^{(\alpha)}$ are the elements of $t_k$. We now define $\hat T_4$ by.
\begin{equation}
\hat T_4(\theta)=\frac{y_3(\theta)}{\hat T_2(\theta)}.
\end{equation}
Using the Y-system equations and (\ref{etas}) it is easy to show that
the roots of $\hat T_4$ are the set $t_4$ and it satisfies (\ref{fund})
with $k=4$. Similarly we construct $\hat T_5, \hat T_6,\dots$ recursively.
We also define
\begin{equation}
\hat T_1(\theta)=\frac{y_2(\theta)}{\hat T_3(\theta)}
\end{equation}
and see that its roots are in $t_1$ and it satisfies
\begin{equation}
\hat T_1(\theta+iq)\,\hat T_1(\theta-iq)=Y_1(\theta)\,Y_0(\theta).
\label{fund1}
\end{equation}
So far we have constructed $\hat T_1, \hat T_2,\dots$ satisfying
\begin{equation}
\hat T_k(\theta+iq)\,\hat T_k(\theta-iq)=Y_k(\theta)\,\left[1+\delta_{k1}
\,y_0(\theta)\right]
\end{equation}
for $k\geq1$,
\begin{equation}
\hat T_{k-1}(\theta)\,\hat T_{k+1}(\theta)=y_k(\theta)
\end{equation}
for $k\geq2$ and the T-system equations
\begin{equation}
\hat T_k(\theta+iq)\,\hat T_k(\theta-iq)=1+\hat T_{k+1}(\theta)\,
\hat T_{k-1}(\theta)
\label{hatT}
\end{equation}
for $k\geq2$. The T-system functions can be completed by $\hat T_0$
by requiring (\ref{hatT}) to hold for $k=1$ also. Note that $\hat T_0$ 
does not have any roots in the physical strip.

The last step is to perform a gauge transformation with $g(\theta)=
\varphi_0(\theta+iq)$. This gives
\begin{equation}
\phi(\theta)=\varphi_0(\theta)\,\varphi_0(\theta+2iq)
\end{equation}
and
\begin{equation}
T_k(\theta)=\varphi_0(\theta+i(k+1)q)\,\varphi_0(\theta-i(k+1)q)\,
\hat T_k(\theta).
\end{equation}
Note that $T_k$ has the same set of roots in the physical strip ($t_k$)
as $\hat T_k$ and it is a smooth deformation of the correspnding T-function
in the large $\ell$ Bethe Ansatz solution.

Having constructed the T-system elements we now turn to the TQ-relations
(\ref{TQ}). It is possible to show that they are equivalent to the single
second order difference equation \cite{KLWZ}
\begin{equation}
\phi(\theta)\,Q(\theta-2iq)+\phi(\theta-2iq)\,Q(\theta+2iq)=A(\theta)\,
Q(\theta),
\label{diff}
\end{equation}
where
\begin{equation}
A(\theta)=\frac{\phi(\theta)\,T_0(\theta-3iq)+\phi(\theta-2iq)\,T_2(\theta-iq)}
{T_1(\theta-2iq)}.
\end{equation}
In the large $\ell$ limit this can be further simplified: 
\begin{equation}
A(\theta)=T_1(\theta).
\end{equation}

Throughout this paper our main assumption was that the exact solution
is a smooth deformation of the one obtained from the Bethe Ansatz
at large $\ell$. Here we make this statement more precise. Since the
exact TBA equations differ from the ones valid in the large $\ell$ limit
only by terms related to $Y_0$, which is exponentially small in the
physical strip, we obviously have
\begin{equation}
Y_k(\theta)\sim Y^{\rm BA}_k(\theta),\qquad\quad \vert{\rm Im\,}\theta\vert
<\frac{\pi}{2},\qquad k=1,2,\dots,
\end{equation}
where $\sim$ here means up to exponentially small corrections (in $\ell$)
and we introduced the superscript $^{\rm BA}$ for objects of the Bethe Ansatz
solution. Next, from the set of Y-system equations (\ref{Y}) we see that
the deformation is actually exponentially small in the larger domain
\begin{equation}
Y_k(\theta)\sim Y^{\rm BA}_k(\theta),\qquad\quad \vert{\rm Im\,}\theta\vert
<\frac{k\pi}{2},\qquad k=1,2,\dots.
\end{equation}
Inspecting the relation between the T-system and Y-system functions we
can see that
\begin{equation}
T_k(\theta)\sim T^{\rm BA}_k(\theta),\qquad\quad \vert{\rm Im\,}\theta\vert
<\frac{(k+1)\pi}{2},\qquad k=1,2,\dots.
\end{equation}
Finally from the TQ-relations (\ref{TQ}) we see that it is natural to 
assume that
\begin{equation}
Q(\theta)\sim Q^{\rm BA}(\theta)=Q_0(\theta),\qquad\quad 
\vert{\rm Im\,}\theta\vert>0.
\end{equation}
Thus $Q(\theta)$ is close to the polynomial $Q_0(\theta)$ in the upper
half plane (and its complex conjugate $\bar Q(\theta)$ in the lower
half plane).

In the large $\ell$ limit it can be shown \cite{PS} that the two 
linearly independent
solutions of (\ref{diff}) are $Q_0(\theta)$ (which is a polynomial of 
degree $M$) and $R_0(\theta)$, an other polynomial of degree $2N+1-M>M$.
We will assume a similar polynomial behaviour of $Q(\theta)$ in the
upper half plane for large $\theta$ thus we can characterize the smoothly 
deformed solution $Q(\theta)$ uniquely by the requirement that 
(in the upper half plane)
\begin{equation}
Q(\theta)\sim\theta^M,
\end{equation}
asymptotically for large $\theta$.

\section{NLIE in Fourier space}

In this Appendix we derive the NLIE in the special case where the
functions $T_1$ and $T_2$ have only real roots in the physical strip
and also the roots of $Q(\theta)$ satisfy
\begin{equation}
\vert{\rm Im}\,u_j\vert<\gamma+\frac{\pi}{2},\qquad
Q(u_j)=0,\quad j=1,2,...M.
\label{Qroots}
\end{equation}
The set of real $T_1$ roots will be denoted by $\{r_j\}$. This is the
union of the set of physical rapidities with the set of type I holes. 
This important special case covers all 2-particle states discussed
in this paper (at least for large enough volume) and many other
multi-particle states of interest. The derivation in the most
general case is more complicated, but goes essentially along the
same lines.

We start with some definitions. To any bounded meromorphic function $\psi(z)$
we associate the Fourier transform of its logarithmic derivative along a line
parallel to the real axis:
\begin{equation}
\psi(z)\quad\Longrightarrow\quad \tilde\psi(k,\alpha)=
\int_{-\infty}^\infty{\rm d}x\,{\rm e}^{ikx}\,\frac{\psi^\prime(x+i\alpha)}
{\psi(x+i\alpha)}.
\end{equation}
For the complex conjugate we have:
\begin{equation}
\bar\psi(z)\quad\Longrightarrow\quad \tilde{\bar\psi}(k,\alpha)=
\tilde\psi^*(-k,-\alpha).
\end{equation}
The function $\psi(z)$ has roots at $R_\mu$ and poles at $P_\nu$. We have
\begin{equation}
{\rm e}^{-k\alpha}\tilde\psi(k,\alpha)
-{\rm e}^{-k\beta}\tilde\psi(k,\beta)=2\pi i\left\{
\sum_{\alpha<{\rm Im}R_\mu<\beta}{\rm e}^{ikR_\mu}
-\sum_{\alpha<{\rm Im}P_\nu<\beta}{\rm e}^{ikP_\nu}\right\}.
\label{REL1}
\end{equation}
In the limit $\beta\to+\infty$ we have
\begin{equation}
{\rm e}^{-k\alpha}\tilde\psi(k,\alpha)=2\pi i\left\{
\sum_{{\rm Im}R_\mu>\alpha}{\rm e}^{ikR_\mu}
-\sum_{{\rm Im}P_\nu>\alpha}{\rm e}^{ikP_\nu}\right\}.
\label{REL2}
\end{equation}

We start from (\ref{bk},\ref{Bk}) and (\ref{bbbar},\ref{BBbar}), which we
recall here for $k=1$:
\begin{eqnarray}
b_1(\theta) &=& \frac{Q(\theta+3i q)}{\bar{Q}(\theta-3i  
q)}
\frac{T_1(\theta-iq)}{\phi(\theta+iq)},  
\label{b1}  \\
B_1(\theta)&=&\frac{Q(\theta+iq)}{\bar{Q}(\theta-3iq)}
\frac{T_2(\theta)}{\phi(\theta+iq)}.   \label{B1}
\end{eqnarray}
\begin{eqnarray}
b_1(\theta) \, \bar{b}_1(\theta)&=&\ell_1(\theta)
\,=\,\left[1+y_1(\theta)\right]\,\left[
1+{\rm e}^{-\ell{\rm cosh}\theta}y_1(\theta)\right],
\label{bb1} \\
B_1(\theta+iq) \, \bar{B}_1(\theta-iq)&=&\ell_2(\theta)
\,=\,1+y_2(\theta).
\label{BB1}
\end{eqnarray} 
We now define
\begin{equation}
\tau_1(k)=2\pi i\sum_j{\rm e}^{ikr_j},\qquad
\tau_2(k)=2\pi i\sum_j{\rm e}^{ikh_j}.
\end{equation}
Some further definitions:
\begin{equation}
\tau(k)={\rm e}^{-qk}\tilde T_1(k,q)-\tilde T_2(k,q),
\end{equation}
\begin{equation}
\tilde\ell_1(k)=\tilde\ell_1(k,0),\,\,\quad
\tilde\ell_2(k)=\tilde\ell_2(k,0),\,\,\quad
\tilde b(k)=\tilde b_1(k,\gamma),\,\,\quad
\tilde B(k)=\tilde B_1(k,\gamma).
\end{equation}

>From the ratio of (\ref{b1}) and (\ref{B1}) we have
\begin{equation}
\tilde b(k)=\tilde B(k)+\tilde T_1(k,\gamma-q)+\tilde Q(k,\gamma+3q)-
\tilde T_2(k,\gamma)-\tilde Q(k,\gamma+q).
\end{equation}
Since $Q(\theta)$ is analytic in the upper half plane and
has no roots in $[\gamma+q,\infty]$ we conclude using
(\ref{REL2}) that
\begin{equation}
\tilde Q(k,\gamma+q)=\tilde Q(k,\gamma+3q)=0,\qquad k>0.
\end{equation}
Furthermore using (\ref{REL1}) we see that
\begin{equation}
\tilde T_1(k,\gamma-q)={\rm e}^{k(\gamma-2q)}\tilde T_1(k,q)
+{\rm e}^{(\gamma-q)k}\tau_1(k),\qquad
\tilde T_2(k,\gamma)={\rm e}^{k(\gamma-q)}\tilde T_2(k,q)
\end{equation}
and thus we can write
\begin{equation}
\tilde b(k)=\tilde B(k)+{\rm e}^{(\gamma-q)k}\left[
\tau(k)+\tau_1(k)\right],\qquad k>0.
\label{bpos}
\end{equation}

We now use the facts that

\noindent $T_1(\theta)$ has no roots in $[-q,-q+\gamma]$,

\noindent $Q(\theta)$ has no roots in $[3q,3q+\gamma]$,

\noindent $\phi(\theta)$ has no roots in $[q,q+\gamma]$,

\noindent $\bar Q(\theta)$ has no roots in $[-3q,-3q+\gamma]$,

\noindent and conclude from (\ref{b1}) using (\ref{REL1}) that
\begin{equation}
\tilde b_1(k,0)={\rm e}^{-k\gamma}\tilde b(k)
\end{equation}
and using (\ref{bb1}) that
\begin{equation}
\tilde b_1(k,0)+\tilde b_1^*(-k,0)=\tilde \ell_1(k)=
{\rm e}^{-k\gamma}\tilde b(k)+{\rm e}^{k\gamma}\tilde b^*(-k).
\end{equation}
This can be used to write a relation also for negative $k$:
\begin{equation}
\tilde b(k)={\rm e}^{k\gamma}\tilde\ell_1(k)-{\rm e}^{2k\gamma}\tilde B^*(-k)
-{\rm e}^{k(\gamma+q)}\left[\tau^*(-k)-\tau_1(k)\right],\quad k<0.
\label{bneg}
\end{equation}

Similarly using that

\noindent $T_2(\theta)$ has no roots in $[\gamma,q]$,

\noindent $Q(\theta)$ has no roots in $[\gamma+q,2q]$,

\noindent $\phi(\theta)$ has no roots in $[\gamma+q,2q]$,

\noindent $\bar Q(\theta)$ has no roots in $[\gamma-3q,-2q]$,

\noindent we conclude from (\ref{B1}) using (\ref{REL1}) that
\begin{equation}
\tilde B_1(k,q)={\rm e}^{k(q-\gamma)}\tilde B(k)
\end{equation}
and also using (\ref{BB1}) we have
\begin{equation}
\tilde\ell_2(k)={\rm e}^{k(q-\gamma)}\tilde B(k)+{\rm e}^{k(\gamma-q)}
\tilde B^*(-k).
\label{l2B}
\end{equation}
We now go back to the original relations (\ref{YT}) and (\ref{modYT})
and write
\begin{equation}
\tilde\ell_1(k)=\tilde T_1(k,q)+\tilde T_1(k,-q)-
\tilde\phi(k,q)-\tilde{\bar\phi}(k,-q)
\label{tl1}
\end{equation}
and
\begin{equation}
\tilde\ell_2(k)=\tilde T_2(k,q)+\tilde T_2(k,-q)-
\tilde\phi(k,2q)-\tilde{\bar\phi}(k,-2q).
\label{tl2}
\end{equation}
Since $\bar\phi(\theta)$ has no roots in $[-2q,-q]$ and $\phi(\theta)$
has no roots in $[q,\infty]$ we see that
\begin{equation}
{\rm e}^{2qk}\tilde{\bar\phi}(k,-2q)={\rm e}^{qk}\tilde{\bar\phi}(k,-q)
\end{equation}
and
\begin{equation}
\tilde\phi(k,2q)=\tilde\phi(k,q)=0,\qquad k>0.
\end{equation}
Further we note that
\begin{equation}
{\rm e}^{kq}\tilde T_i(k,-q)-{\rm e}^{-qk}\tilde T_i(k,q)=\tau_i(k),\quad
i=1,2.
\end{equation}
Combining (\ref{tl1}) and (\ref{tl2}) we thus have
\begin{equation}
{\rm e}^{-qk}\tilde\ell_1(k)-\tilde\ell_2(k)=\tau(k)\left(1+{\rm e}^{-2qk}
\right)+{\rm e}^{-2qk}\tau_1(k)-{\rm e}^{-qk}\tau_2(k),
\quad k>0,
\end{equation}
and, after some algebra and using (\ref{l2B})
\begin{equation}
\begin{split}
\tau(k)=g(k)\Big\{\tilde\ell_1(k)-{\rm e}^{k(2q-\gamma)}\tilde B(k)
&-{\rm e}^{k\gamma}\tilde B^*(-k)\\
&+\tau_2(k)-{\rm e}^{-qk}\tau_1(k)\Big\},
\quad k>0,
\end{split}
\label{taupos}
\end{equation}
where
\begin{equation}
g(k)=\frac{1}{2{\rm cosh}\, qk}.
\end{equation}
>From this we get for negative $k$, after complex conjugation:
\begin{equation}
\begin{split}
\tau^*(-k)=\tau_1(k)
+g(k)\Big\{\tilde\ell_1(k)-{\rm e}^{-k\gamma}\tilde B(k)
&-{\rm e}^{k(\gamma-2q)}\tilde B^*(-k) \\
&-\tau_2(k)-{\rm e}^{-qk}\tau_1(k)\Big\},
\quad k<0.
\end{split}
\label{tauneg}
\end{equation}
Finally we use (\ref{taupos}) in (\ref{bpos}) and similarly
(\ref{tauneg}) in (\ref{bneg}) and get for positive $k$:
\begin{equation}
\begin{split}
\tilde b(k)=g(k)\Big\{{\rm e}^{-qk}\tilde B(k)&-{\rm e}^{k(2\gamma-q)}
\tilde B^*(-k)\\
&+{\rm e}^{k(\gamma-q)}\tilde\ell_1(k)+{\rm e}^{k(\gamma-q)}\tau_2(k)+
{\rm e}^{k\gamma}\tau_1(k)\Big\},\quad k>0
\end{split}
\end{equation}
and for negative $k$:
\begin{equation}
\begin{split}
\tilde b(k)=g(k)\Big\{{\rm e}^{qk}\tilde B(k)&-{\rm e}^{k(2\gamma+q)}
\tilde B^*(-k)\\
&+{\rm e}^{k(\gamma-q)}\tilde\ell_1(k)+{\rm e}^{k(\gamma+q)}\tau_2(k)+
{\rm e}^{k\gamma}\tau_1(k)\Big\},\quad k<0.
\end{split}
\end{equation}
Combining the last two equations we now write the NLIE in Fourier space as
\begin{equation}
\begin{split}
\tilde b(k)=g(k)\Big\{{\rm e}^{-q\vert k\vert}
\tilde B(k)&-{\rm e}^{2k\gamma-q\vert k\vert)}
\tilde B^*(-k)\\
&+{\rm e}^{k(\gamma-q)}\tilde\ell_1(k)+{\rm e}^{k\gamma-q\vert k\vert)}
\tau_2(k)+
{\rm e}^{k\gamma}\tau_1(k)\Big\}.
\end{split}
\end{equation}
This is the Fourier space version of the first equation of the NLIE 
(\ref{nlie1}) in our special case.

\end{document}